\documentclass[aps,twocolumn,superscriptaddress,pra,showpacs]{revtex4}
\usepackage{epsfig}
\usepackage{amsmath}

\newcommand{\ket}[1]{|#1 \rangle}

\begin{document}

\title{Scalable Error Correction in Distributed Ion Trap Computers}
    \author{Daniel K. L. Oi}
    \email{D.K.L.Oi@damtp.cam.ac.uk}
    \affiliation{Centre for Quantum Computation, Department of Applied
      Mathematics and Theoretical Physics, University of Cambridge, Wilberforce
      Road, Cambridge CB3 0WA,
    United Kingdom}
    \author{Simon J. Devitt}
     \affiliation{Centre for Quantum Computation, Department of Applied
      Mathematics and Theoretical Physics, University of Cambridge, Wilberforce
      Road, Cambridge CB3 0WA,
    United Kingdom}
    \affiliation{Centre for Quantum Computing Technology, Department of Physics,
    University of Melbourne, Victoria, Australia}
    \author{Lloyd C.L. Hollenberg}
    \affiliation{Centre for Quantum Computing Technology, Department of Physics,
    University of Melbourne, Victoria, Australia}
    

\begin{abstract}
  A major challenge for quantum computation in ion trap systems is scalable
  integration of error correction and fault tolerance. We analyze a distributed
  architecture with rapid high fidelity local control within nodes and
  entangled links between nodes alleviating long-distance transport. We
  demonstrate fault-tolerant operator measurements which are used for error
  correction and non-local gates. This scheme is readily applied to linear ion
  traps which cannot be scaled up beyond a few ions per individual trap but
  which have access to a probabilistic entanglement mechanism. A
  proof-of-concept system is presented which is within the reach of current
  experiment.
\end{abstract}


\maketitle

\section{Introduction}

Quantum Computation (QC)~\cite{Nielsen} poses extreme challenges for coherent
control of large-scale quantum systems. Coping with decoherence and imperfect
control implies the use the of error correction codes (QEC) and fault tolerant
operation. However, maximizing the threshold for arbitrary computation
requires the ability to perform multiple, simultaneous operations between
qubits and minimal communication and transport overheads. Incorporating these
features in a scalable manner is a major goal for all potential system
implementations~\cite{KMW2002,MTCCC2005,Taylor2005,HGFW2005,DGH2005}.

A promising candidate for quantum information processing is the ion trap with
superb coherence and few qubit control having been already
demonstrated~\cite{LeibfriedEA2005,BlattEA2005}. However, it is difficult to
effectively control more than a few tens of ions in a single trap, hence
several ideas have been proposed to overcome this limitation. Multiple
microtraps can be constructed in the same structure with ions shuttled between
them (CCD architecture)~\cite{KMW2002}. The main disadvantage of a CCD trap is
the difficulty in designing a micro-trap structure that allows for maximum
parallelizability for both inter- and intra-logical operations. Shuttling
heats ions up, requiring additional cooling in the interaction regions and
slowing operation. Additionally, large numbers of electrodes and lasers would
be required in a single device~\cite{Steane2004}.

Alternatively, ions in separate trap structures may be made to interact via a
photonic
bus~\cite{CZKM1997,Pellizari1997,SL2000,KLHLW2003,DBMM2004,DMMMKM2006,vEKCZ1999}.
If used directly to implement two-qubit gates, photon loss from the bus is a
major problem and requires additional QEC overhead~\cite{RHG2005,RRM2006}.
Alternatively, the photonic bus can mediate the generation of entanglement
between traps which can then be used to perform
gates~\cite{DB2003}~\footnote{Cluster state generation for One-Way
  Computation~\cite{RB2001} becomes straightforward in this system, local
  control in each node effectively allows linear scaling~\cite{BBFM2005}.
  Failure of an entangling operation does not destroy any prior links, and
  purification procedures can be used to increase the fidelity of the final
  cluster state.  In this paper however, we will concentrate upon conventional
  gate-based QC.}.

The use of entanglement for intra-computer communication is not a new idea.
For example, this has been proposed in CCD ion trap designs where EPR pairs
are created locally and then the halves sent to entanglement stations
distributed among a sea of qubits~\cite{MTCCC2005}. These entangled pairs
would then be used to teleport qubits between memory storage and processing
regions, circumventing the problem of directly transporting data across the
whole computer. However, since the entangled pairs themselves are created
locally and then the separate halves physically moved to where they are
required, this neccessitates the use of quantum repeaters and extensive
purification.  Furthermore in~\cite{MTCCC2005}, all 49 qubits of a second
level encoded logical qubit are teleported requiring many EPR pairs for
transport in both directions.

\begin{figure}
\includegraphics[width=0.45\textwidth]{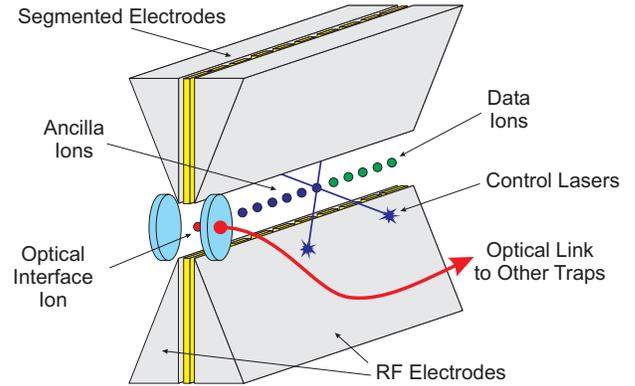}
\caption{An ion trap node. A single ion trap contains enough ions for a single
  encoded qubit, ancillas and an interface ion. Conventional single and
  two-qubit operations are performed via axial phonon modes of the
  trap. The interface ion may be entangled with its counterpart in another
  identical trap via photon interference and path erasure. The resultant Bell
  link is used to perform inter-trap operations. The simplest node consists of
  a single optically coupled ion, five ions encoding a single logical qubit,
  and several ancilla for fault tolerant operations and singlet purification.}
\label{fig:trap}
\end{figure}

In this paper, we outline an proof-of-concept architecture based around an ion
trap processing node containing a relatively small number of ions representing
an encoded first level error corrected qubit, ancillas for fault-tolerant
operation, and an interface ion which can be entangled with its counterpart in
another node (Fig.~\ref{fig:trap}). An abstract basis of the scheme was
suggested in~\cite{DB2003}, but here we analyze a concrete realization, taking
particular attention to the requirements of error correction and fault-tolerant
operation. In particular, we show how local and non-local logical operations
can be reliably performed directly between two nodes via \emph{operator
  measurements}, from which scaling to an arbitrary sized quantum computer
follows. A small prototype is presented which is within reach of current
experiment.

The paper is laid out as follows: The basic architecture is covered in
Section~\ref{sec:architecture}, the use of operator measurements to implement
gates is in Section~\ref{sec:operation}, the preparation of encoded Bell states
is in Section~\ref{sec:qec}, fault-tolerant implementation of non-local
operators is in Section~\ref{sec:faulttolerant}, architecture scale-up is in
Section~\ref{sec:extending}, optimizing node design is in
Section~\ref{sec:node}, and concluding remarks in Section~\ref{sec:conclusion}.

\section{Architecture}
\label{sec:architecture}

\begin{figure}
\includegraphics[width=0.45\textwidth]{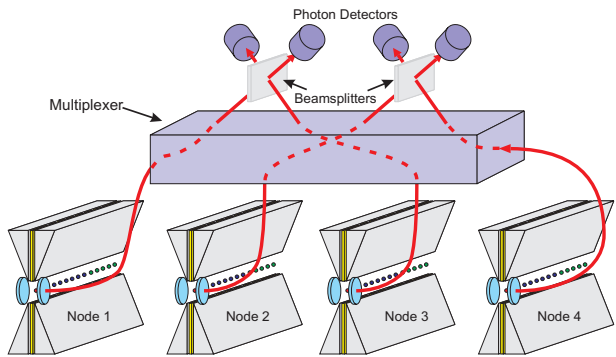}
\caption{Distributed Ion Trap Architecture. The whole computer consists of a
  set of identical nodes, each holding a few physical qubits encoding a logical
  qubit, and associated ancillas. The nodes are connected by optical fibre
  linking the interface ions in each node. A heralded probablistic procedure
  entangles pairs of ions in separate nodes via interference and path erasure.
  An optical multiplexer allows arbitrary pairs of nodes to be entangled, and
  parallel operation is achieved using multiple beam-splitters and detectors.}
\label{fig:scheme}
\end{figure}

The basic architecture is illustrated in Fig.~\ref{fig:scheme}. A network of
local processing nodes are connected by optical fibres and a multiplexing
switch. In each trap node is a small array of ions
upon which conventional single and intra-trap
two-qubit operations can be performed. Pairs of nodes can be optically linked
to beamsplitters and single photon detectors which entangle the interface qubits
when subjected to appropriate laser excitation and conditioned upon a correct
sequence of detector clicks. The resulting Bell pair is then be used to perform
inter-node operations.

\subsection{Operation}
\label{sec:operation}

We start off with all qubits initialized. Intra-trap operations are used to
prepare encoded qubits. We assume that each trap can hold a sufficient number
of ions to encode a logical qubit plus an appropriate number of ancilla
ions for error correction in at least the first level of concatenation.
Single qubit, non-trivial, logical operations (for example the
$T$ gate~\cite{Fowler}) are performed with the assistance of ancilla qubits in
the local trap. For inter-node two-qubit logical operations, instead of
directly interacting data qubits via the photonic bus, we
instead create Bell pairs spanning the nodes. By local operations and classical
communication (LOCC), two-qubit gates can be performed without risking data
loss between nodes.

\subsubsection{Inter-node operations and encoded Bell state preparation}
\label{sec:qec}

\begin{figure}
\includegraphics[width=0.45\textwidth]{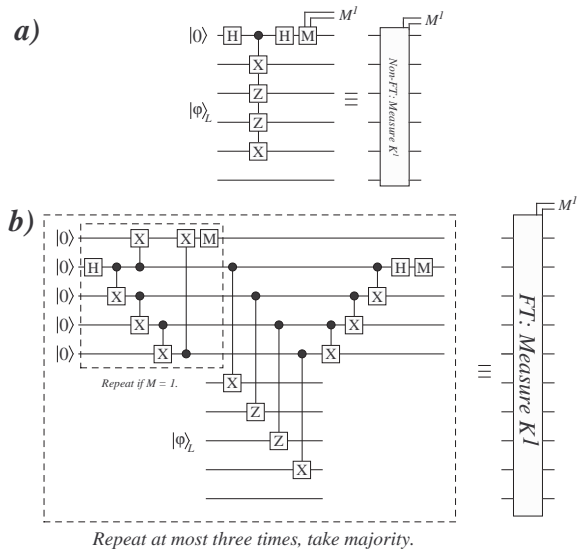}
\caption{Quantum circuit measuring the stabiliser $K^1$ for the [[5,1,3]]
  quantum code. a) non-Fault-Tolerant circuit b) basic Fault-Tolerant circuit.
  The fault-tolerant circuit first requires the preparation and verification
  of a four qubit GHZ state.  If the verification measurement $=1$, then the
  ancilla block is reset and prepared again.  To protect against $Z$ errors in
  the ancilla block, the circuit is repeated up to three times and a majority
  vote of the syndrome results is taken.}
\label{fig:cor}
\end{figure}

As an example of inter-node operations, consider the preparation of a logically
encoded Bell state between two separate nodes. Each node houses between seven
and fourteen ions depending on whether fault-tolerant error correction and gate
operations are employed.  The data ions in each trap will be encoded using the
[[5,1,3]] code~\cite{five1,shor1}, which is the smallest full quantum code,
requiring five ions for a single logically encoded qubit protected from at most
one error. The stabiliser structure~\cite{Gottesman} for the [[5,1,3]] code,
$K^{i}$ $i \in \{1,2,3,4\}$, and the logical bit ($\bar{X}$) and phase
($\bar{Z}$) operations are specified by,
\begin{eqnarray}
K^{1}=XZZXI,&\ &K^{2}=IXZZX,\nonumber \\
K^{3}=XIXZZ,&\ &K^{4}=ZXIXZ,\nonumber \\
\bar{X}=XXXXX,&\ &\bar{Z}=ZZZZZ.
\label{eq:stab}
\end{eqnarray}
Where $X$ and $Z$ are the Pauli $\sigma_x$ and $\sigma_z$ operators, $I$ is the
2$\times$ 2 identity matrix, and the tensor product is implied.  Error
correction using stabiliser codes is straightforward \cite{shor1,Nielsen}, each
of the four generators $K^{i}$ are measured [Fig. \ref{fig:cor}(a)] either
sequentially using a single ancilla, or simultaneously using four ancilla.
Each of the sixteen possible four-bit results represent one of the correctable
single qubit errors, as well as the case where no error occured. At a minimum,
fault-tolerant measurement of the stabilisers requires a four qubit GHZ state
as an ancilla block [Fig. \ref{fig:cor}(b)].  Additionally, a fifth qubit is
used to verify the GHZ state against possible $X$ errors which can subsequently
propagate to the data block.  Therefore the minimum number of ions in a single
trap needed for logical encoding and correction is six, while a total of ten
ions are needed to employ full fault-tolerant correction sequentially.

The interaction between logical qubits in separate nodes is mediated by
interface ions entangled into Bell pairs by any one of a number of
methods~\cite{CCGZ1999,PHBK1999,BKPV1999,BPH2003}.  It has been shown that some
two-qubit gates can be performed using Bell pairs via
LOCC~\cite{Gottesman1998,VC2004,DBMM2004,DMMMKM2006}. A large class of quantum
codes, known as Calderbank-Shor-Steane (CSS) codes, allow logical
controlled-$\sigma_x$ (CNOT) and controlled-$\sigma_z$ (CZ) gates to be applied
\emph{block-wise} between two data blocks, which are also inherently
fault-tolerant.  However, the [[5,1,3]] quantum code is \emph{not} a CSS code
and block-wise CNOT or CZ gates are not possible. This also means that the more
rapid method of error correction introduced by Steane~\cite{steane}, requiring
a larger ancilla block, will not work with the [[5,1,3]] code.  However, a
logical CNOT or CZ interaction between two logical blocks of data can be
performed for \emph{any} code that allows for block-wise single-qubit bit
and/or phase operations using fewer interface qubits than the standard
block-wise approach.  This method, first proposed in \cite{leung} and
\cite{devitt} uses the same basic element as error correction, namely operator
measurements.

A CZ gate between two qubits can be written in terms of operators on an
arbitrary two qubit state $\ket{\psi}$ as,
\begin{equation}
CZ\ket{\psi} = \frac{1}{2}(II +ZI+IZ-ZZ)\ket{\psi}.
\end{equation}
To achieve this transformation on an arbitrary two-qubit state, we append an
ancilla qubit prepared in the state $\ket{+}=(\ket{0}+\ket{1})/\sqrt{2}$,
and measure the operators $ZIZ$ and $IZX$ over the three qubit system.
After these measurements, and assuming that the qubits are always projected to
a $+1$ eigenstate of these operators (otherwise local corrections can be
applied), the final state is given by
\begin{equation}
\frac{1}{2}(III + ZIZ + IZX + ZZ(X.Z))\ket{\psi}\ket{+}.
\end{equation}
Since the Pauli operators $X$ and $Z$ anti-commute and that
$X\ket{+}=\ket{+}$ and $Z\ket{+}=\ket{-}$, the state can be
re-written as,
\begin{equation}
  \frac{1}{2}((II+IZ)\ket{\psi}\ket{+}+(ZI-ZZ)\ket{\psi}\ket{-}),
\end{equation}
after which the ancilla is then measured in the computational basis.  If the
measurement result is $\ket{0}$, $\ket{\psi}$ is projected to CZ$\ket{\psi}$,
otherwise it is projected to $(IZ).\text{CZ}\ket{\psi}$ upon which a local $IZ$
correction is then applied.

We use the above method to perform a logical CZ across two nodes. A single
physical Bell state is prepared between two nodes each containing a logical
qubit. Each half of the Bell state is used as a control qubit on the respective
data block of an encoded qubit. For example, to measure the logical
$\bar{Z}\bar{Z}$ operator across two logical blocks, local CZ gates are applied
between each Bell pair qubit and the five ions representing the single logical
qubit in each node. For a general state of two logical qubits $\ket{\psi}_L$,
the transformation is
\begin{equation}
\frac{1}{\sqrt{2}}(\ket{00}+\ket{11})\ket{\psi}_L\rightarrow
\frac{1}{\sqrt{2}}(\ket{00}II+\ket{11}\bar{Z}\bar{Z})\ket{\psi}_L,
\end{equation}
where $\bar{Z}$ is as in Eq.\ref{eq:stab}, a logical phase gate for the
[[5,1,3]] code. A local Hadamard gate is applied to both interface qubits,
leading to the state,
\begin{equation}
\frac{1}{2\sqrt{2}}\sum_{j,k=0}^1
\ket{jk}(II+(-1)^{j+k}\bar{Z}\bar{Z})\ket{\psi}_L.
\end{equation}
Measuring the parity of interface qubits projects the data qubits into a $\pm
1$ eigenstate of $\bar{Z}\bar{Z}$ for an even/odd parity result, hence
performing the required measurement.

Measuring an appropriate sequence of operators will enact a \emph{logical}
controlled phase rotation across two nodes. To perform the full CZ gate, an
ancilla is needed which is finally measured in the computational basis.
This ancilla \emph{does not} need to be a fully encoded logical qubit in
its own trap, it can just as easily be a single ion contained in either the
control or target trap.  However, to maintain fault-tolerance this ancilla
qubit \emph{should} be encoded. By using operator measurements between traps.
inter-logical operations can be performed directly on the [[5,1,3]] encoded
data using only one interface qubit per trap.

Localizing a single logical qubit plus appropriate ancilla ions for local error
correction has several advantages. Intra-trap operations have been demonstrated
on up to eight ions~\cite{BlattEA2005} so local operations and error correction
should suffer minimal overhead. Probabilistic entanglement
generation~\cite{CCGZ1999,PHBK1999,BKPV1999,BPH2003} does not pose a problem
for inter-node operations as local error correction can preserve coherence
between the generation of entangled links. This is substantially more
advantageous than other highly distributed schemes~\cite{DBMM2004} where
\emph{all} ions interact via these non-local linkages which would be highly
susceptible to memory errors while waiting for non-local links to be
established. If required, local control permits the use of purification
protocols in order to increase inter-trap gate fidelity~\cite{DB2003}. Also, by
structuring all inter-logical operations such that they are mediated by
entangled links, larger trap structures, for example the CCD design of
Kielpinski \emph{et. al.}~\cite{KMW2002}, need only be designed, and
optimized, for local error correction.

From an experimental standpoint, the preparation of a logically encoded Bell
pair does not require the operator CZ gate in full.  If the initial state is
$\ket{0}_L\ket{0}_L$, the measurement of the operator
$\bar{X}\bar{X}$ is sufficient to produce the state
$(\ket{0}_L\ket{0}_L+\ket{1}_L\ket{1}_L)/\sqrt{2}$.  Hence experimental
demonstration of encoded Bell state preparation does not need the third
ancilla qubit required by the operator CZ gate.

\begin{figure}
\includegraphics[width=0.45\textwidth]{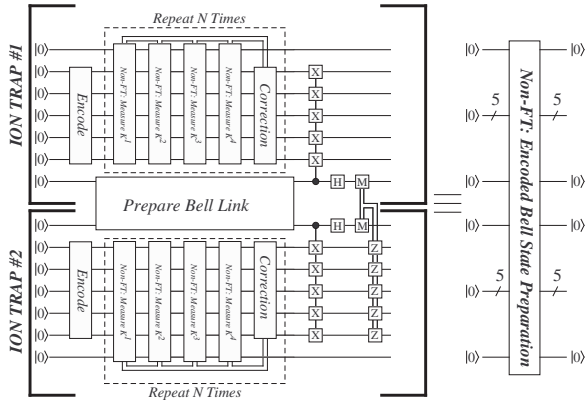}
\caption{Full quantum circuit for non-fault-tolerant, non-local preparation of an
  encoded Bell state across two nodes.  Five ions in each trap are first
  encoded into the $\ket{0}_L$ state after which local error correction is
  repeated continuously (say $N$ times) to protect against memory errors while
  the Bell link is created. Once the Bell link is created, each interface ion
  is used as a control qubit for a blockwise $X$ gate on each trap.  The
  interface qubits are then measured locally and a classical $\bar{Z}$ gate is
  applied to the second trap if the measurement result has odd parity.  The
  final state of the two traps is the encoded Bell state,
  $(\ket{0}_L\ket{0}_L+\ket{1}_L\ket{1}_L)/\sqrt{2}$.}
\label{fig:NFT}
\end{figure}

Without maintaining fault-tolerance, we only require five (data block) $+$ one
(ancilla) $+$ one (interface) ions per trap to implement the protocol.
Fig.~\ref{fig:NFT} shows the complete quantum circuit required to implement the
state preparation non-fault-tolerantly, assuming that $N$ full local error correction cycles are
performed in the time required to prepare the inter-trap Bell link.

\subsubsection{Fault-Tolerant encoded state preparation.}
\label{sec:faulttolerant}

This general method for preparing a distributed, encoded Bell state is not
fault-tolerant.  Utilizing a single qubit for correction allows errors to
cascade into the data block.  Also, the Bell pair interface can induce
\emph{logical} errors if it is not prepared correctly. 

Maintaining fault-tolerance for local error correction is fairly
straightforward~\cite{shor,Nielsen}.  The stabilizers for the [[5,1,3]] code
have a maximum weight of four, hence the ancilla ion used for correction is
replaced with five ions, four of which are prepared in the entangled state
$(\ket{0000}+\ket{1111})/\sqrt{2}$, after which the fifth is used to verify
the ancilla state against $X$ errors that can propagate to the data block.  If
verification fails, the ancilla block is reset and re-prepared.  Once the
state is verified, each of the four ancilla ions are coupled to the data
block, with local CNOT and Hadamard gates (depending on the stabiliser
structure) and measured to determine the syndrome.  To protect against $Z$
errors, occurring or propagating to the ancilla block, the syndrome is
measured multiple times.  At least two syndrome measurements are made, if they
disagree a third syndrome is measured and a majority vote taken
[Fig.~\ref{fig:cor}(b)].  We adapt this general method to operator measurement
gates between traps.

Errors in the Bell link between node, either during preparation or during
operation, can lead to multiple errors propagating to the data blocks. For
fault-tolerance we again use several ancillas, thereby ensuring that only one
ancilla qubit interacts with one qubit within the data block.  Measuring the
operator $\bar{U}_1 \otimes \bar{U}_2$, where
$\bar{U}_j\in\{\bar{X},\bar{Z}\}$, requires a Bell link between the two nodes
and the ability to perform CNOT or CZ gates between each qubit in the Bell pair
and their respective data block.  If an $X$ error occurs on the Bell pair then
this can propagate to possibly all of the qubits within one of the nodes.  If
this occurs, the single qubit error will induce a \emph{logical} error. To
counter this, we introduce several more ancilla qubits into each node and
verify the interface qubit state before coupling ions to the data block.

\begin{figure}
\includegraphics[width=0.45\textwidth]{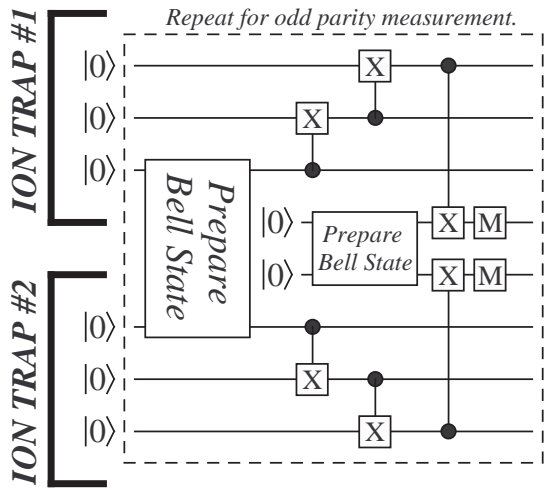}
\caption{Circuit to prepare and verify the interface ancilla
  blocks for fault-tolerant operator measurement on the [[5,1,3]] code.  The
  ancilla state requires the preparation of two Bell links between the separate
  data traps.  After the local CNOT gates the second Bell link is measured.  If
  the measurement result has odd parity, the interface block is reset and
  re-prepared.  Local error correction can be performed on each data block
  while waiting for a verified interface ancilla block.}
\label{fig:FT}
\end{figure}

The required circuit needed to prepare a sufficient interface ancilla for the
[[5,1,3]] code is shown in Fig~\ref{fig:FT}. We use two Bell pairs and two
additional ancilla qubits in each node that are coupled to the original Bell
pair through CNOT gates. After preparation, the ancilla blocks of the two nodes
are in the state,
\begin{equation}
\ket{\text{An}}=
\frac{1}{2}(\ket{000}_1\ket{000}_2+\ket{111}_1\ket{111}_2)(\ket{00}_v+\ket{11}_v)
\end{equation}
The subscripts 1 and 2 represent the three ancilla qubits within each node,
while the subscript $v$ represents a second Bell link between the two nodes
used for ancilla verification. Now a CNOT gate is performed between the last
qubit in each node and the verification Bell state. If no errors have occurred
then the CNOT operations will leave the verification Bell pair invariant.
Considering all the possible single $X$ error locations during the preparation
of the ancilla state we find the following unique states are possible.
\begin{eqnarray}
\ket{\text{An}}&=&
\frac{1}{2}(\ket{000}_1\ket{000}_2+\ket{111}_1\ket{111}_2)
(\ket{01}_v+\ket{10}_v),\nonumber\\
\ket{\text{An}}&=&
\frac{1}{2}(\ket{000}_1\ket{111}_2+\ket{111}_1\ket{000}_2)
(\ket{01}_v+\ket{10}_v),\nonumber\\
\ket{\text{An}}&=&
\frac{1}{2}(\ket{000}_i\ket{011}_j+\ket{111}_i\ket{100}_j)
(\ket{01}_v+\ket{10}_v),\nonumber\\
\ket{\text{An}}&=&
\frac{1}{2}(\ket{000}_i\ket{001}_j+\ket{111}_i\ket{110}_j)
(\ket{01}_v+\ket{10}_v),\nonumber\\
\end{eqnarray}
Where $[i,j] \in \{[1,2],[2,1]\}$. The verification qubits are measured in the
computational basis and if an even parity result is obtained, then the ancilla
state is verified, otherwise either the ancilla or verification qubits have
experienced a single $X$ error and we repeat the preparation.

The inclusion of the second Bell link between two nodes and the additional
ancillas allows the verification of the ancilla state prior to coupling it
to the data qubits, protecting the ancilla state from a single $X$ error.  Phase
errors in the ancilla block result in an incorrect determination regarding
which eigenstate the data qubits are projected to.  To protect against this,
the operator is measured two to three times and a majority vote taken.  At
each stage, error correction can be continuously performed on each data block
while the interface ancilla block is prepared and verified.

In Section~\ref{sec:qec} we showed how the Bell link can allow the measurement
of a given Hermitian operator. This required performing a controlled operation
from the Bell link qubits to each of the data qubits.  For the [[5,1,3]] code,
this allows for a logical $\bar{X}$ and/or $\bar{Z}$ operator measurement since
these logical operations can be performed block-wise.  To maintain
Fault-Tolerance, this would require five ancilla qubits in each node connected
to the Bell pair. However we can reduce this to three in each
node by exploiting the stabiliser structure of the [[5,1,3]] code.

Any given logical state $\ket{\psi}_L$ encoded with the [[5,1,3]] code is
stabilised by the operators $K^1$ to $K^4$. Therefore $\ket{\psi}_L =
K^i\ket{\psi}_L$, $i \in\{1,2,3,4\}$. If a logical $\bar{Z}$ ($\bar{X}$)
operation is performed on the state, it is not necessary to apply five single
qubit $Z$ ($X$) gates, but we can redefine the logical operators in terms of the
stabilisers.  Consider the first stabiliser $K^1 = XZZXI$.  Then,
\begin{equation}
\bar{Z}\ket{\psi}_L=\bar{Z}K^1\ket{\psi}_L=(X.Z)II(X.Z)Z\ket{\psi}
\label{eq:zbar}
\end{equation}
therfore only three operators, hence four ancilla qubits, are required
for the interface block, instead of five.

The total number of ions in each trap for full fault-tolerant local correction
and coupling between the traps is fourteen. Five ancilla
ions are needed for local fault-tolerant error correction of the five ion
logical qubits, while four ions are needed as the interface ancilla block
including two non-local Bell links, one to
actually link the traps and one to verify the ancilla block. The total number
of qubits needed for this scheme and the number of non-local Bell links for a
general quantum code is significantly less depending on the size of the code
used. For a general, CSS, $n$ qubit code correcting a single error, and
assuming that no purification protocols are used, $n$ Bell links are required
to perform a block-wise CZ gate in one time step. In contrast, this scheme only
requires two Bell links (regardless of the code size) to perform a CZ gate in
several time steps. In between each step, local error correction can be
performed to protect against memory errors.

\subsection{Extending the scheme to larger architectures}
\label{sec:extending}

The above scheme of preparing a non-local encoded Bell state between two
separate nodes can easily be extended to a much larger distributed system.
Each node would be designed to house a single logical qubit, or several
logical qubits for a general [[$n$,$k$,$d$]] code with $k >1$.  
If multiple concatenation levels are warranted, then the node system
would also have the requisite number of qubits and routing system to allow
full Fault-Tolerant error correction at all levels. The inter-logical
operations, at the highest level of encoding are then performed using the
non-local Bell links and the operator measurement protocol.  For a multiply
concatenated ($m^{th}$) level qubit, the operator measurement formalism can be
extended in a straightforward manner.

\begin{figure}
\includegraphics[width=0.45\textwidth]{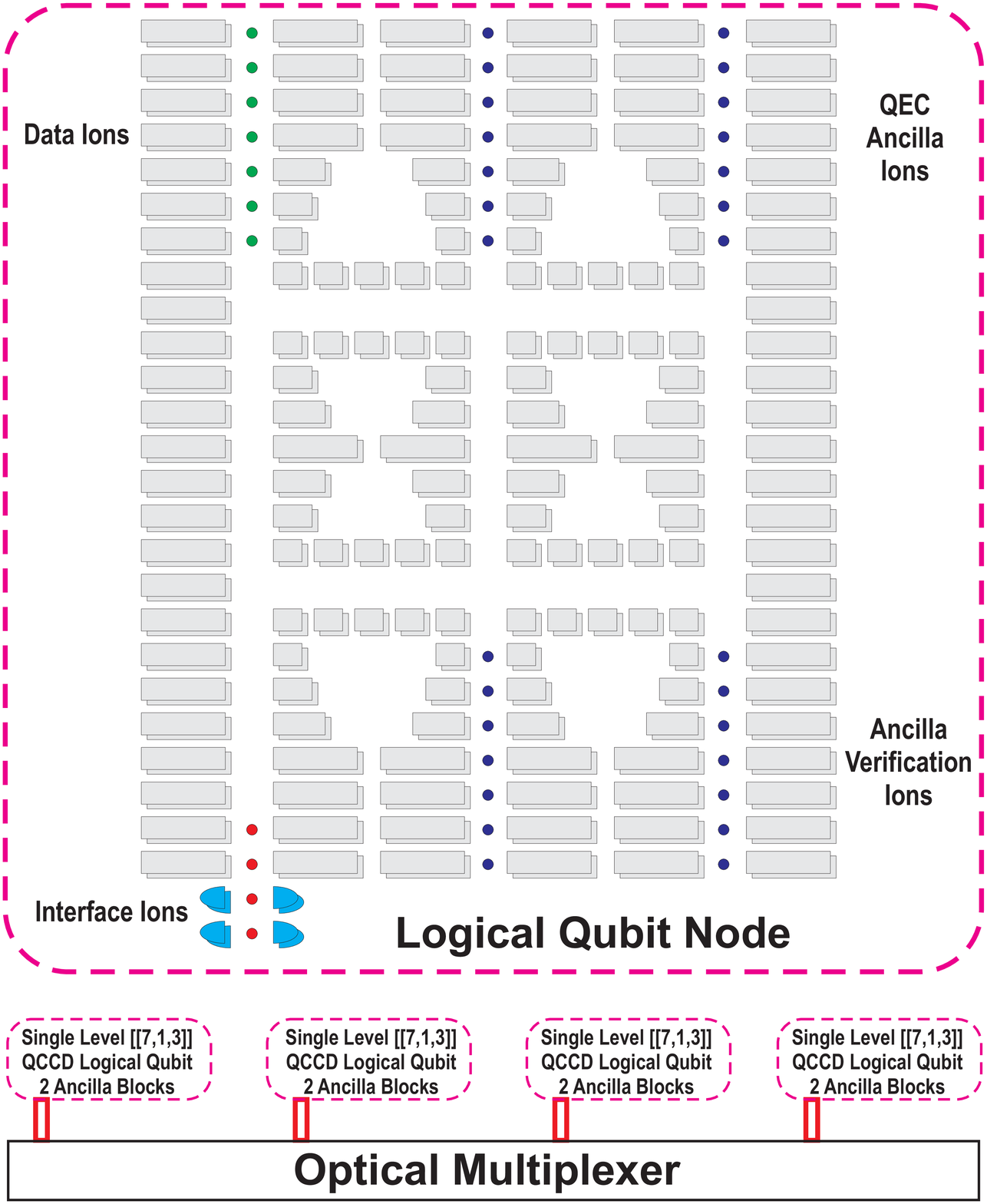}
\caption{CCD micro-trap structure for a single logical qubit using the 
  [[7,1,3]] Steane code.  Each chip houses 39 ions: 7 data ions, 28 ancilla
  ions (allowing for simultaneous preparation and verification of two ancilla
  blocks using Steane's rapid correction method~\cite{steane}) and 4 interface
  ions for coupling to other logical qubits.  The interface state required for
  the [[7,1,3]] code is identical to the [[5,1,3]] code since each of the seven
  dimensional stabilizers for the Steane code has weight four, hence
  $\text{Wt}(\bar{Z}) = \text{Wt}(\bar{X}) = 3$.  Each of these chip nodes can
  then be connected to the optical multiplexer, increasing the total size of
  the quantum computer as needed.}
\label{fig:CCD}
\end{figure}

For all CSS codes, block-wise $Z$ and $X$ operations are possible.  Hence to
measure these operators across two nodes at the $m^{th}$ level of encoding, a
CZ or CNOT gate is performed between each half of a Bell link pair of qubits
and all of the physical qubits in the two separately encoded nodes.
Fault-tolerance would require a similar ancilla system as that used in the
[[5,1,3]] example.  Instead of using two Bell links and four ancilla qubits per
trap, the total number of ancillas will be equal to
\begin{equation}
\text{Number of Ancilla} = \text{Wt}(\bar{U})^m+1,
\end{equation}
where $m$ is the concatenation level and $\text{Wt}(\bar{U})$ is the minimum
weight of the $n$ qubit operation that invokes a blockwise logical $U$
operation on the $k^{th}$ qubit (if multiple qubits are encoded within a single
node).  The number of Bell links required between nodes remains constant at
two, unless a quantum code is employed that encodes multiple logical qubits.
In this case inter-logical operations between states located in different nodes
will require two Bell links for each pair of qubit interactions between nodes.
Figure~\ref{fig:CCD} shows an example structure for a distributed computer
using the CCD trap design.  Each CCD chip is designed exclusively for a single
logical qubit encoded with the [[7,1,3]] Steane code.  Each chip houses seven
data qubits, an additional 28 ancilla qubits which would allow for the
simultaneous preparation and verification of two separate ancilla blocks using
the rapid method of Steane~\cite{steane}, and the four interface qubits which
are required for fault-tolerant operator measurements using the Steane code.
Each chip is manufactured and characterized separately and would be
\emph{plugged} in to the optical multiplexer, linking it to the rest of the
computer.

Within a larger architecture, the logical qubits needed for a given quantum
algorithm are interspersed with logically encoded ancilla traps that are then
used to perform logical CZ gates using the methods described in
Section~\ref{sec:qec}.

\section{Node Design}
\label{sec:node}

To summarize the architecture, each node should satisfy the key requirements:
\begin{itemize}
\item A sufficient number of long lived physical qubits for an error corrected
  logical qubit.
\item An additional number of ancilla qubits for error correction and operator
  measurements. Measurement of these ancilla should be fast and reliable. The
  absolute coherence time of these ancilla may be traded against fast operations.
\item A qubit which can be entangled with its counterpart in another node.
  This process can be probabilistic but heralded.
\item Fast and reliable single and two-qubit operations within the node for
  single logical qubit operations, error correction, operator measurements, and
  entanglement purification.
\end{itemize}
If multiple Bell links are required with a single node, the state of the
interface qubit can be swapped to an ancilla and the interface qubit
re-entangled. Of course, multiple interface qubits would allow for parallel
entangling operations but are not strictly necessary. Entanglement purification
may be required to increase the fidelity of the entangled links between traps.
Nested entanglement pumping~\cite{DBCZ1999} reduces the number of ancilla
required for high fidelity Bell pairs.

Segmentation of a linear trap could be used to isolate the interface ion from
the rest of the trap until required. By suitable geometry, the interface region
would not impinge on intra-trap operation, either by phonon coupling or photon
scattering.  When entanglement is needed, the trap potentials are rearranged so
that an ancilla ion could be placed into a common mode with the interface ion
and quantum state transfer performed, afterwards which the ancilla would be
brought back to the rest of the ions for further processing.

Though we have primarily considered a linear Paul trap as a node, one could
replace it with any other small qubit system as long as the above requirements
are met, e.g. a CCD trap with an optical interface region as in
Section~\ref{sec:extending}. The Penning trap~\cite{COCWST2006} has also been
suggested as a candidate for quantum computation, with hundreds or thousands of
ions in a single two-dimensional Coulomb crystal~\cite{DBHHS1998,PC2006} and
two-qubit gates via transverse phonon modes~\cite{ZMD2006}. The large number of
physical qubits would allow larger code words protecting against multiple
errors and/or optimised for different error models. However, the rotation of
the crystal would complicate ion addressing~\footnote{This could be achieved in
  principle, for instance, with a rotating dove prism synchronised with the
  crystal rotation locked to an external driving frequency~\cite{HBMI1998}.}
and would restrict strong cavity coupling to the central ion~\footnote{In such
  a system, the interface qubit would increasingly be the bottleneck,
  restricting inter-node operation, unless scheduling could restrict operations
  within qubits within a single node whereever possible.}.

Within each node, the physical qubits play different roles opening up the
possibility of optimization of their separate properties. The data qubits
require long coherence times, whilst we may want to optimise the ancilla qubits
for fast operations and measurement. The interface qubit should have suitable
optical properties for the entanglement generation procedure. In a Paul ion
trap, different ionic species could be utilized and loaded in order by
frequency selective ionization~\cite{BBHPEN2006}. The use of heavier ions (such
as $\text{Cd}^+$) for data storage may reduce gate errors due to spontaneous
decay from intermediate metastable states~\cite{WinelandEA2003}, or else direct
microwave driving of hyperfine transitions could eliminate this
entirely~\cite{WB2003}. Lighter ions could be used as ancilla in order to lower
the mass of the ion string and hence raise the axial phonon frequencies aiding
cooling and two-qubit gate times.

To reduce the number of ions in each node, the use of multiple levels in the
ground hyperfine manifold to encode multiple qubits could be
considered~\cite{ARDA}. Since measurement is likely to distinguish the state of
all the encoded qubits of an ion, this method may not be suitable for data
qubits, but is not necessarily a drawback for use for ancilla qubits which en
bloc are measured and initialized repeatedly.

\section{Conclusion}
\label{sec:conclusion}

We have proposed the use of entanglement to directly implement non-local
operations between separately housed logical qubits. These ideas may also be
applicable to other physical quantum computing implementations which satisify
the requirements in Section~\ref{sec:node}~\footnote{For example, ion traps
  connected via superconducting elements~\cite{TRBZ2004}, superconducting
  qubits connected via microstriplines and polar molecules~\cite{RDDLSZ2006,
    ADDLMRSZ2006} or solid state spin qubits coupled via superconducting
  cavity QED~\cite{BI2006}.}. The entanglement is created by a point-to-point
process which reduces routing difficulties and enables parallel operation.
Logical operations via operator measurements require minimal entangled
resources compared to a directly teleported sequence of block-wise gates but
still retains fault tolerance. By keeping data local to a single node, the
node can be of comparatively simple design and size, optimized for local high
fidelity operations. The technique should be able to be generalized to
multi-qubit operations utilizing multi-partite entangled states and may serve
as the basis for a full scalable quantum computing architecture. A proof of
principle demonstration with two traps containing seven or eight ions and an
optical interface each is within the reach of current
experiment~\cite{KLHLW2003,BlattEA2005}. Even simpler to demonstrate are
operator measurement gates, the optical interface could be omitted and a gate
performed between two three-qubit encoded (single X or Z-error) logical qubits
coupled via a single Bell pair. Such a distributed architecture is a strong
alternative to monolithic designs.

\begin{acknowledgments}
  DKLO acknowledges the support of the Cambridge-MIT Institute Quantum
  Information Initiative, EU grants RESQ (IST-2001-37559) and TOPQIP
  (IST-2001-39215), EPSRC QIP IRC (UK), and Sidney Sussex College, Cambridge.
  SJD acknowledges the support of the Rae \& Edith Bennett Travelling
  Scholarship. LH and SJD are supported in part by the Australian Research
  Council, the Australian Government and the US National Security Agency
  (NSA), Advanced Research and Development Activity (ARDA), and the Army
  Research Office (ARO) under contract number W911NF-04-1-0290. The authors
  would also like to thank D. Segal for enlightening discussions.
\end{acknowledgments}

\end{document}